\documentclass[global,twocolumn]{svjour}
\usepackage{latexsym}
\usepackage{amsmath}
\usepackage{graphics}
\usepackage{cite}

\journalname{}
\begin{document}
\title{Effects of flake size on mode-locking behavior for flake-graphene saturable absorber mirrors}

\author{M.I. Hussain \inst{1}\and B.V. Cunning \inst{2}\and L.S. Booth \inst{1} \and M.J. Petrasiunas \inst{1}\and C.L. Brown \inst{3}\and D. Kielpinski \inst{1}
}
                   
\institute{Center for Quantum Dynamics, Griffith University, Nathan, QLD 4111, Australia \and International Institute for Carbon-Neutral Energy Research (WPI-I2CNER), Kyushu University, Nishi-ku, Fukuoka 819-0395, Japan\and Queensland Micro and Nanotechnology Center, Griffith University, Nathan, QLD 4111, Australia\\
\email{mahmood.hussain2@griffithuni.edu.au}
}

\date{Received: date / Revised version: date}
%
\maketitle
\begin{abstract}
After advent of graphene as a saturable absorber many experiments have been conducted to produce short pulse duration pulses. Here, we have measured the properties of flake-graphene saturable absorber mirrors of various flake sizes dependent on fabrication technique. These mirrors enabled us to obtain a large mode-locking bandwidth of 16nm in an erbium-doped fiber laser. Mirrors with large flake size and multi-layered thickness induce strong pulse shaping and reflect mode-locked train of pulses with very large bandwidths.
\end{abstract}
\section{Introduction}
\label{intro}
Ultrafast pulses are playing a key character in modern research now a days \cite{fermann2002}. In the class of ultrafast pulses, femtosecond pulses are eminent because of their extremely short time duration. Owing to this property of femtosecond pulses whole field of laser matter interaction has worn a new look \cite{dieeter2000, zhang2010}. We are now able to steer and monitor quantum processes in real time \cite{moore2012}. In general, their applications list span from medical to industry \cite{mgower2000, dausinger2004, marion1999}. As a result of their substantial role, their production also stands fundamental. Various techniques have been proposed for producing mode-locked ultra-short pulses \cite{french1995, park2011}. Among these, saturable absorption of light on a laser cavity mirror is often used for mode-locking ultrafast fiber lasers, which basically exploit the principle of intensity dependent absorption of laser pulses \cite{Fermann2009}. This scheme has been studied with various materials, including semiconductor saturable absorber mirror (SESAM) in which semiconductor heterostructures are formed by low and high band-gap materials \cite{keller1992, keller1996, Okhotnikov2004}. Soliton based fiber laser mode-locking has also been studied in single wall carbon nanotubes (SWCNT) \cite{zsun2012, marteniz2013}. Beside SESAM and SWCNT, graphene thin films (mono or multi atomic layer) have been investigated in the past for mode-locking ultra-short laser pulses \cite{ zhang2009,  bao2009, sun2010, sunn2010, song2010, tan2010, bo2011, kim2011, bao2011, xu2011, huang2012, xu2012}. Similarly, graphene film deposition on reflective substrate has also been used as one of the fiber laser cavity mirror, known as graphene saturable absorber mirror (GSAM).
Low loss flake GSAM has been reported to produce ultrafast pulses with low gain fiber laser \cite{ben2011}. Flake graphene saturable absorber with polymer has been used in transmission mode to achieve wide optical bandwidth \cite{popa2010}. Comparative studies have been done to characterise the mono, few and multi-layered atomic graphene for mode-locking, fabricated by mechanical exfoliation method \cite{marteniz2011}. GSAM was used to mode-locked fiber laser at very high repetition rate of 10GHz \cite{martinez2012}. In the recent past, single atomic layer of graphene has been used to mode-lock fiber laser operating at 2$\mu$m wavelength \cite{lagatsky2013}.

\begin{figure}

\resizebox{0.48\textwidth}{!}{%
  \includegraphics{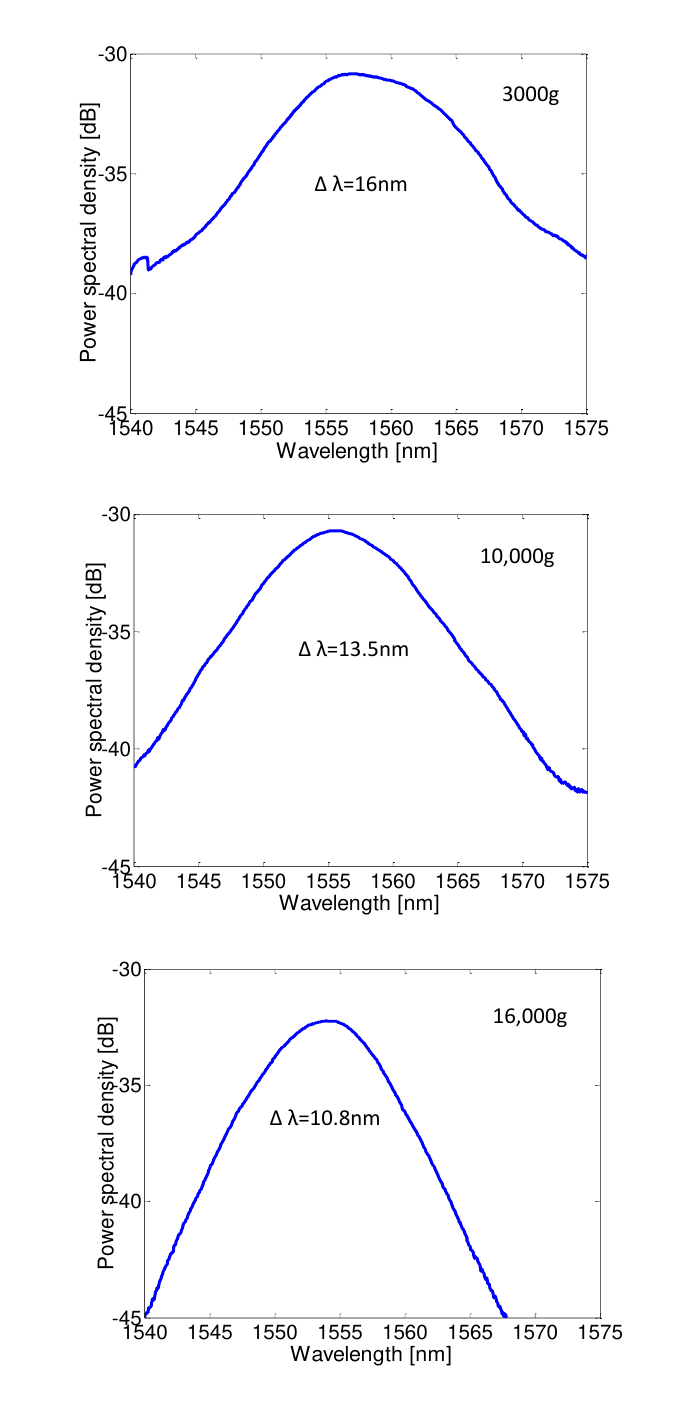}
}

\caption{Optical spectrums and bandwidths at FWHM for each sample.}
\label{bandwidth}       
\end{figure}

Here, we have investigated the wide mode-locked-bandwidth (ML-BW) and saturable absorption characteristics achievable through various GSAMs of different flake size and thickness. Flake size and thickness differs on the basis of graphene thin film fabrication technique, which has been discussed in the next section. Flake size varies roughly from 120nm to 450nm. We found that multi-layered large graphene flakes with lateral size of order of 450nm can produce ML-BW more than 16nm. Best results we achieved are shown in Fig. \ref{bandwidth}, which depicts the optimized optical spectrum with full-width at half-maximum (FWHM) numbers. We were able to achieve a stable and maximum spectral ML-BW of 16nm. Upto our best knowledge, it is the largest bandwidth ever reported in the literature with any GSAM and 28\% wider than previously reported in reflection mode \cite{ben2011}.
\section{Fabrication of graphene saturable absorbers}
A number of GSAMs were fabricated by transferring films consisting of few-layer graphene flakes to gold mirrors. Suspensions of graphene flakes were initially prepared by the sonication of flake graphite. Typically, flake graphite (5g) was bath sonicated (150W) for four days in 500mL of a 1\% aqueous solution of Triton X-100 surfactant. Subsequently, the suspension was centrifuged at either 3,000g, 10,000g or 16,000g, three times for 90 minutes, collecting the top 90\% of the supernatant for the subsequent centrifugation, creating stable dispersions of flake graphene. The varying centrifugation speed allowed for control of the lateral dimensions of the graphene flakes \cite{ben2014}. Thin films of flake graphene were prepared by slow vacuum filtration of an aliquot of the graphene solution through a 50nm pore size mixed cellulose ester/nitrocellulose membrane filter. The aliquot size was adjusted to create graphene films of small, medium and large arbitrary thicknesses on top of the membrane filter. The films were transferred to the mirror substrate by first by pre-wetting the mirror with isopropyl alcohol and clamping the film (face down) to the mirror. After the isopropyl alcohol had evaporated, the membrane filter was removed by successive dissolutions in acetone. The mirror substrate consisted of a 160nm SiO$_2$ layer atop of a thin film of reflective gold to ensure the graphene film lay at an antinode of the reflected laser light.

\begin{figure}

\resizebox{0.48\textwidth}{!}{%
  \includegraphics{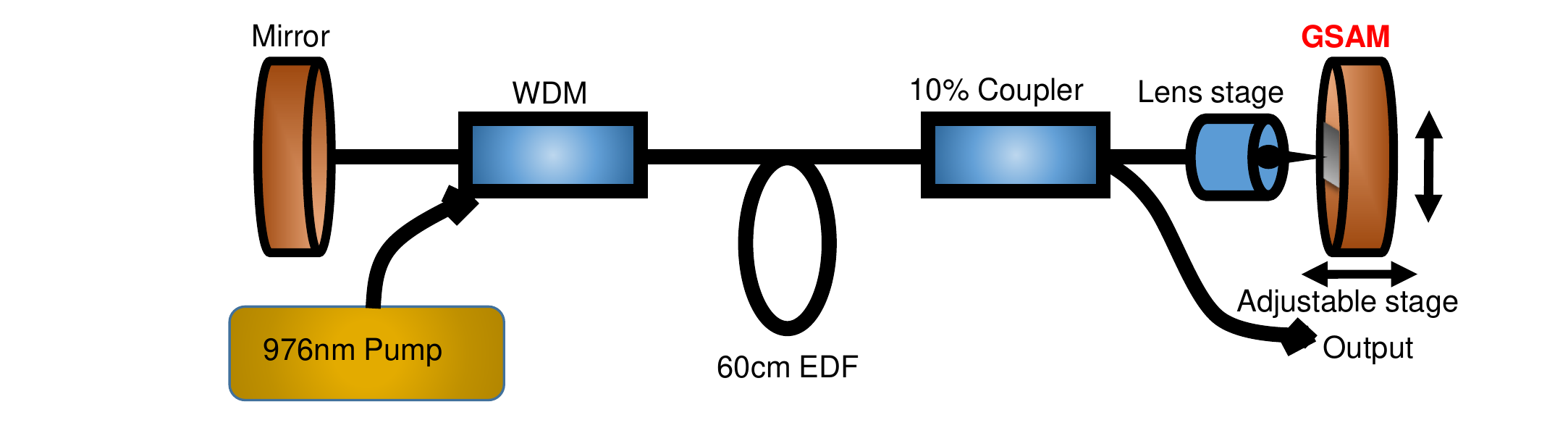}
}

\caption{Schematic of setup for dispersion managed fiber laser mode-locking. EDF--Erbium-doped fiber, WDM--wavelength-division multiplexer, GSAM--graphene saturable absorber mirror.}
\label{setup}       
\end{figure}

\begin{table}
\caption{Maximum mode-locked bandwidths, ratio of saturated to unsaturated ($\alpha_{sat}/\alpha_{unsat}$) loss and approximate lateral flake size for different GSAMs.}
\begin{tabular}{cccc}
\hline\hline
Sample & Flake size & Bandwidth (Max) & $\alpha_{sat}/\alpha_{unsat}$\\
\hline
3,000g & 445nm & 16nm & 0.93$\pm$0.03 \\
10,000g & 185nm & 13.5nm & 0.86$\pm$0.06\\
16,000g & 120nm & 10.8nm & 0.80$\pm$0.02 \\
\hline
\end{tabular}
\label{tab} 
\end{table}

\section{Results and discussion}
\label{sec:2}
\begin{figure}

\resizebox{0.48\textwidth}{!}{%
  \includegraphics{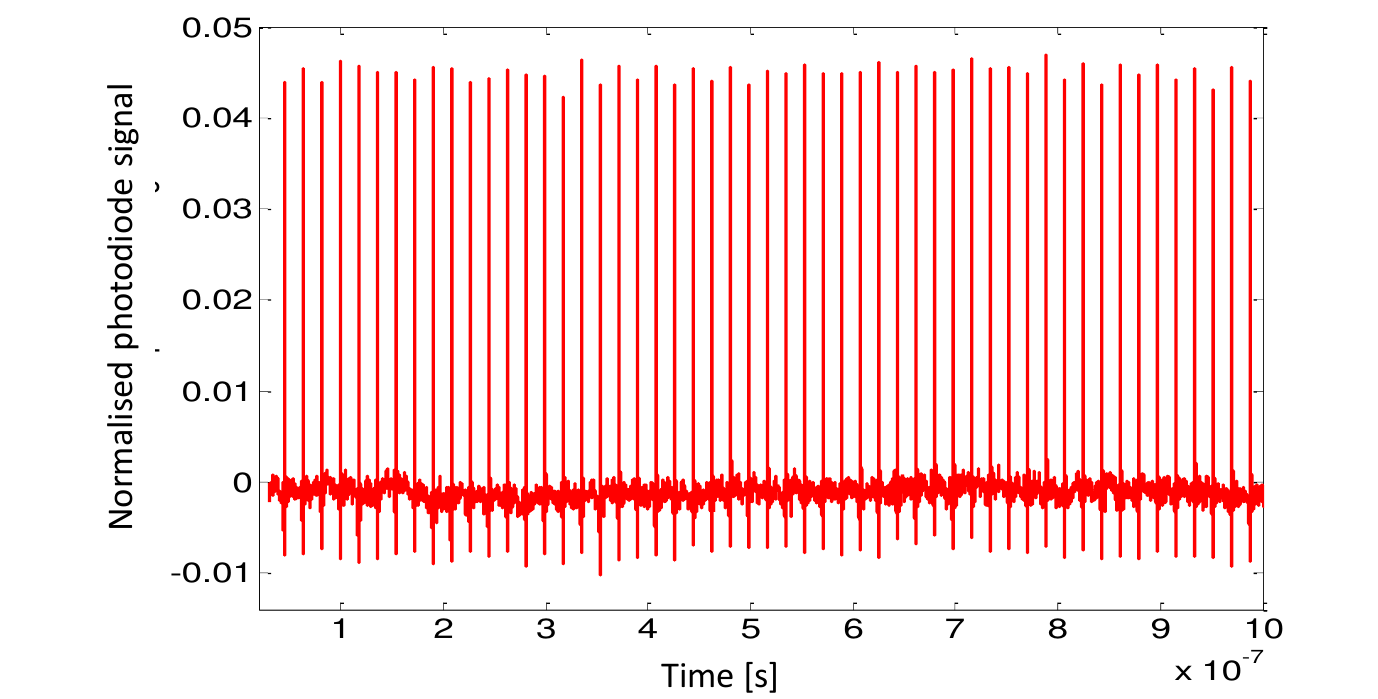}
}
\caption{Photo diode trace of mode-locked pulse train coming out of laser cavity in time domain, shows amplitude variation in pluses is less than 5\%.}
\label{pulsetrain}    
\end{figure}

Different samples of GSAM were employed to mode-lock erbium-doped fiber laser (EDFL) at different bandwidths. EDFL used is a standard linear cavity fiber laser with average power 0.8mW and operating wavelength 1550nm, Fig. \ref{setup} shows schematic of our setup.  Total dispersion in the cavity for lengths $\approx$(Er80 = 46cm, SMF28 = 136cm) is about -920fs$^2$. We used fiber coupled Faraday mirror at one end and on other end GSAM samples were used in free space. For all cases the repetition rate was same $\approx$ 57MHz because each sample was placed at the focus of 2.8mm lens in free space, Fig. \ref{pulsetrain} shows the photodiode trace of the mode-locked pulses in time domain. Group velocity dispersion (GVD) was overcome by engineering the lengths of the single mode fiber and erbium-doped gain fiber. Pump power was always greater than 50mW for each sample to initiate mode-locking. ML-BW spectrum was mainly optimised by using polarization controllers for all cases. Table \ref{tab} shows the results for bandwidth and saturable loss ratio after optimizing the laser for bandwidth. Samples prepared at 3,000g were mode-locked very quickly and remained stable for many hours as compared to those prepared at high centrifuge speeds.

\begin{figure}

\resizebox{0.48\textwidth}{!}{%
  \includegraphics{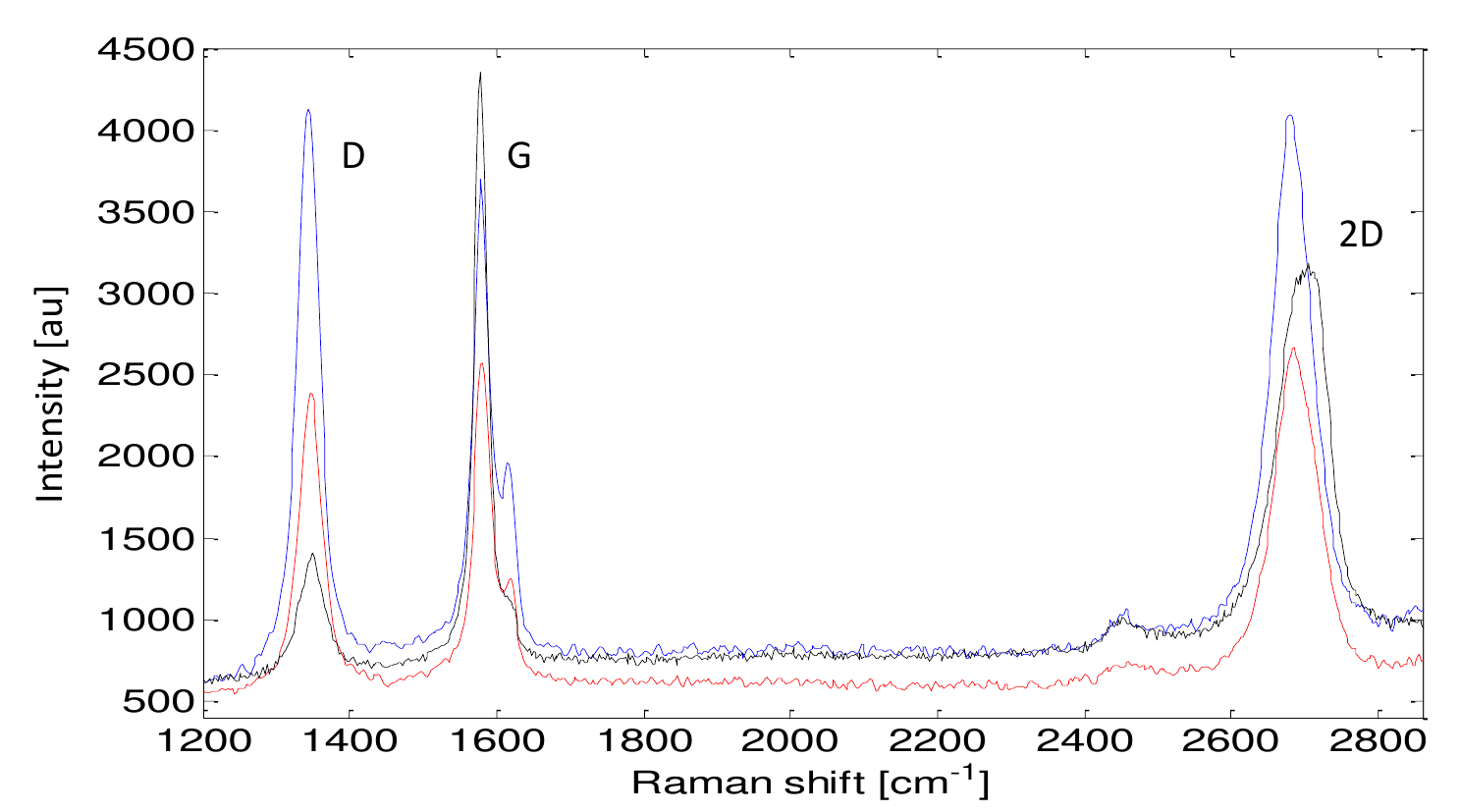}
}

\caption{Raman spectroscopy of each sample which shows D, G, 2D modes, black--3000g, red--10000g, blue--16000g. Data from \cite{ben2014}.}
\label{raman}       
\end{figure}

The Raman spectroscopy of GSAMs is shown in Fig. \ref{raman}. The D band originates due to graphene edges and its intensity shows defect content in the samples \cite{gupta2009}, which increases with increase in the centrifuge speed. The Raman spot size is of the order of 1 micron, so for the higher centrifugation speeds there is a very large concentration of edges under the laser spot in Raman spectroscopy. Hence samples prepared at higher centrifuge speed have more edge defect concentration and small flake size than low centrifuge speed. As observed in \cite{ben2014}, the spectrum shows that there are few basal plane defects, and the D band arises from the flake edges. The shape of the 2D mode of Raman spectra indicates number of layers per flake, with the 2D band splitting into two components. One component is centered at $\approx$ 2670cm$^{-1}$ which represent low layer flake while other is centered at $\approx$ 2700cm$^{-1}$ and is associated with multilayer flakes. From these spectra, we infer that with increase in the centrifugation speed, the fraction of multilayer flakes is decreasing as the heavier and multi-layered flakes sediment out.
Dynamic light scattering technique was used to determine the hydrodynamic size distribution of flakes, as explicitly discussed in the paper \cite{ben2014}. The distributions found in \cite{ben2014} were not well resolved. 

The quantification of ratio of saturated to unsaturated loss ($\alpha_{sat}$/$\alpha_{unsat}$) was done by measuring slope efficiencies. Slope efficiencies were measured during laser operation in mode-locked ($\eta_{ML}$), continuous wave ($\eta_{CW}$) and by reflecting light from gold substrate surface ($\eta_{gold}$). Using these numbers ratio of saturated to unsaturated loss was calculated by using formula \cite{ben2011}
\begin{equation}  \label{ratio}
\dfrac{\alpha_{sat}}{\alpha_{unsat}}=\dfrac{\eta^{-1}_{ML}-\eta^{-1}_{gold}}{\eta^{-1}_{CW}-\eta^{-1}_{gold}},
\end{equation}
 and found that it was $\geq$80\% for our GSAMs, as shown in Table \ref{tab}. Wide ML-BW had been observed by using GSAM prepared at low centrifuge speed, 3,000g, which had large flake size and multi-layered graphene. High layer density increased $\alpha_{sat}$, while $\alpha_{unsat}$ was reduced due to large flake size which decreased the edge concentration and scattering losses. As a result, overall the ratio of $\alpha_{sat}$ and $\alpha_{unsat}$ was high. Samples prepared at high centrifuge speed (10,000g and 16,000g) had small and few layer flakes which made the ratio of $\alpha_{sat}$ and $\alpha_{unsat}$ low. We momentarily observed bandwidths greater than 16nm with 3,000g sample. It is very likely that high centrifuge speed made the graphene flakes smaller and few layer. As a consequence, a higher fraction of the loss is saturable for smaller flakes and it made the mode-locking difficult and shrank the bandwidth. It is quite possible that a lot of particles were not saturated owing to inhomogeneity of particle size and orientation. Loss was still there, even when we dropped down to few layer flakes, because few layer flakes were smaller and had more edge defect concentration which enhanced the scattering losses and hence unsaturated absorption. We also measured the damage threshold of our samples, fluence should be $\leq$500$\mu$J/cm$^2$ for safe operation.
\section{Conclusion}
\label{sec:3}
Graphene films with different lateral flake sizes have been successfully used for mode-locking a fiber laser. The films with the largest diameter flakes mode locked readily and resulted in an optical bandwidth 16nm, the largest bandwidth reported for any graphene based saturable absorber as one of the cavity reflective mirror. GSAM with small flake size and less number of graphene layers produces unstable mode-locking. $\alpha_{sat}$/$\alpha_{unsat}$ and ML-BW increases with increase in the flake size and layer density, which is roughly in conformity with the previous studies \cite{huang2012, marteniz2011}. Smaller flakes have high edge defect concentration which contributes to scattering losses, enhances overall absorption. Table 1 seems to show that smaller flakes have less unsaturable loss as a fraction of total loss. Larger the flake size less would be the $\alpha_{sat}$/$\alpha_{unsat}$ and vice versa.
We believe that better control of the GSAM fabrication process, and thus of the flake size and thickness can provide us optimised flake size that can further increase bandwidth with less losses.

\section{acknowledgments}

This work was funded by the Australian Research Council (ARC) under DP130101613. D.K. was supported by an ARC Future Fellowship (FT110100513).

\end{document}